\documentclass[twocolumn,]{aastex631}

\usepackage{newtxtext,newtxmath}

\usepackage[T1]{fontenc}

\DeclareRobustCommand{\VAN}[3]{#2}
\let\VANthebibliography\thebibliography
\def\thebibliography{\DeclareRobustCommand{\VAN}[3]{##3}\VANthebibliography}

\usepackage{graphicx}	
\usepackage{amsmath}	
\usepackage{multirow}

\newcommand{\rev}[1]{#1}
\newcommand{\sam}[1]{\textcolor{purple}{ SP:}}
\newcommand{\ie}{\emph{i.e.},~}
\newcommand{\eg}{\emph{e.g.},~}

\shorttitle{FRBs May Be Closer Than They Appear}
\shortauthors{Orr et al.}
\begin{document}
\title{Objects May Be Closer Than They Appear: Significant Host Galaxy Dispersion Measures of Fast Radio Bursts in Zoom-in Simulations}


\author[0000-0003-1053-3081]{Matthew E. Orr}
\affiliation{Center for Computational Astrophysics\\ Flatiron Institute\\ 162 Fifth Avenue, New York, NY 10010, USA}
\affiliation{Department of Physics and Astronomy\\ Rutgers University\\ 136 Frelinghuysen Road, Piscataway, NJ 08854, USA}

\author[0000-0001-5817-5944]{Blakesley Burkhart}
\affiliation{Department of Physics and Astronomy\\ Rutgers University\\ 136 Frelinghuysen Road, Piscataway, NJ 08854, USA}
\affiliation{Center for Computational Astrophysics\\ Flatiron Institute\\ 162 Fifth Avenue, New York, NY 10010, USA}

\author[0000-0002-1568-7461]{Wenbin Lu}
\affiliation{Department of Astronomy, University of California at Berkeley, Berkeley, CA 94720, USA}

\author[0000-0002-7484-2695]{Sam B. Ponnada}
\affiliation{TAPIR, California Institute of Technology, Pasadena, CA 91125, USA}

\author[0000-0002-3817-8133]{Cameron B. Hummels}
\affiliation{TAPIR, California Institute of Technology, Pasadena, CA 91125, USA}





\begin{abstract}
We investigate the contribution of host galaxies to the overall Dispersion Measures (DMs) for Fast Radio Bursts (FRBs) using the Feedback in Realistic Environments (FIRE-2) cosmological zoom-in simulation suite.  We calculate DMs from every star particle in the simulated L* galaxies by ray-tracing through their multi-phase interstellar medium (ISM), summing the line-of-sight free thermal electron column for all gas elements within $\pm$20 kpc of the galactic mid-plane.  At $z=0$, we find average (median) host-galaxy DMs of 74 (43) and 210 (94) pc cm$^{-3}$ for older ($\gtrsim$10 Myr) and younger ($\lesssim$10 Myr) stellar populations, respectively.  Inclination raises the median DM measured for older populations ($\gtrsim$10 Myr) in the simulations by a factor of $\sim$2, but generally does not affect the younger stars deeply embedded in H{\small II} regions except in extreme edge-on cases (inclination $\gtrsim 85^\circ$). In kinematically disturbed snapshots ($z = 1$ in FIRE), the average (median) host-galaxy DMs are higher: 80 (107) and 266 (795) pc cm$^{-3}$ for older ($\gtrsim$10 Myr) and younger ($\lesssim$10 Myr) stellar populations, respectively. FIRE galaxies tend to have higher DM values than cosmological simulations such as IllustrisTNG\rev{, with larger tails in their distributions to high DMs}.  As a result,
FRB host galaxies may be closer (lower redshift) than previously inferred.
Furthermore, constraining host-galaxy DM distributions may help significantly constrain FRB progenitor models.


\end{abstract}




\section{Introduction}

First discovered by  \citet{2007Sci...318..777L}, Fast Radio Bursts (FRBs)  are a mysterious class of millisecond duration radio transients that are likely extragalactic in origin.  To date, hundreds of FRBs have been identified and a few tens have localized redshifts through the combined efforts of multiple teams using ASKAP, Parks, MOST, CHIME, DSA-110, and FAST telescopes \citep[\eg][]{2013Sci...341...53T, 2017MNRAS.468.3746C, 2018MNRAS.475.1427B, 2018Natur.562..386S, 2019Natur.566..230C, 2020ApJ...903..152H, 2021ApJ...909L...8N, 2021arXiv210604352T, DSA110}.  Additionally, some FRBs are known to be repeaters, providing some clues about their origins
\citep[\eg][]{2016Natur.531..202S, 2019Natur.566..230C, 2019ApJ...885L..24C, 2020ApJ...891L...6F, 2021ApJ...910L..18B}. Despite this observational progress, the origins of FRBs remain mysterious \citep[][]{2019ARA&A..57..417C, 2019A&ARv..27....4P,2019PhR...821....1P, 2020Natur.587...45Z,
2021arXiv210710113P}\footnote{So much so that popular culture has picked up on the scientific debate, \eg see: \url{https://xkcd.com/2886/}}.  One promising model for FRBs is that they are associated with magnetars; an idea which is strengthened by the detection of an FRB associated with the Milky Way (MW) magnetar SGR 1935+2154 \citep[][]{2020Natur.587...59B, 2020Natur.587...54C, 2020MNRAS.498.1397L, 2020ApJ...899L..27M}. However, given its low energy compared to other FRBs, other source objects are certainly still viable \citep[][]{2019ARA&A..57..417C, 2019A&ARv..27....4P,2019PhR...821....1P, 2020Natur.587...45Z}.  Ultimately, to determine the origins and nature of FRBs, their host galaxy and local environmental conditions must be understood, which requires more precise localization and understanding of the signal characteristics \citep{2023arXiv230316775L,2023MNRAS.518..539M}.

The FRB millisecond signals propagate along the line of sight and undergo 
dispersion and scattering induced by the local Galactic interstellar medium (ISM), the host galaxy ISM, and intergalactic medium (IGM)  \citep[][]{2019A&ARv..27....4P,2019ARA&A..57..417C,2023arXiv230316775L}.  In particular, the dispersion measure (DM) of  FRBs has a wide distribution ranging from $\sim$100 to several thousand $\rm{pc}\, \rm{cm}^{-3}$. The DM is defined as:
\begin{equation}
    {\rm DM}=\int{n_e}{\rm d}l \; ,
\label{eqn:dm}
\end{equation}
where $n_e$ is the number density of free electrons along the line-of-sight and d$l$ is the integral element of line-of-sight. 

If FRBs are indeed extragalactic in origin, the total DM contribution can be written as

\begin{equation}
    \rm{DM_{obs} = DM_{MW,ISM} + DM_{MW,Halo} + DM_{IGM} + DM_{Host}/(1+z)} \; .
\label{eqn:DM_obs}
\end{equation}
As the observed DMs for most FRBs are larger
 then the estimated contribution from the MW ISM ($\rm{DM}_{\rm{MW, ISM}}$) and the MW halo ($\rm{DM}_{MW,Halo}$), most of the DM comes either from the IGM ($\rm{DM}_{IGM}$) or from the host galaxy ISM and halo ($\rm{DM}_{Host}$). 

 \begin{table*}\caption{FIRE-2 Galaxy Physical Properties at Integer Redshift $z = 2-0$}\label{table:gal_props}
   \hskip-2.5cm \begin{tabular}{lccccccccccccc} 
\hline
&  \multicolumn{4}{c}{$z\approx 2$} & \multicolumn{4}{c}{$z \approx 1$} & \multicolumn{4}{c}{$z \approx 0$} \\
\hline
Name & $\log\left(\frac{M_\star}{{\rm M_\odot}}\right)$ & $\log\left(\frac{M_{\rm gas}}{{\rm M_\odot}}\right)$ & $\rm f_{gas}$ & $\frac{\dot M_\star}{\rm M_\odot/yr}$\textdagger
& $\log\left(\frac{M_\star}{{\rm M_\odot}}\right)$ & $\log\left(\frac{M_{\rm gas}}{{\rm M_\odot}}\right)$ & $\rm f_{gas}$  & $\frac{\dot M_\star}{\rm M_\odot/yr}$\textdagger
& $\log\left(\frac{M_\star}{{\rm M_\odot}}\right)$ & $\log\left(\frac{M_{\rm gas}}{{\rm M_\odot}}\right)$ & $\rm f_{gas}$  & $\frac{\dot M_\star}{\rm M_\odot/yr}$\textdagger\\ 
\hline
m12b 
 & 9.83 & 10.11 & 0.65 & \rev{6.7} & 10.51 & 9.95 & 0.22 & \rev{0.10} & 10.96 & 10.30 & 0.18 & \rev{8.8}\\
m12c 
 & 9.28 & 9.78 & 0.76 & \rev{0.37} & 10.18 & 10.16 & 0.49 & \rev{4.2} & 10.80 & 10.29 & 0.24 & \rev{7.4}\\
m12f 
 & 9.94 & 10.20 & 0.64 & \rev{9.8} & 10.42 & 10.02 & 0.28 & \rev{6.7} & 10.92 & 10.30 & 0.19 & \rev{10.7}\\
m12m 
 & 9.51 & 10.04 & 0.76 & \rev{2.13} & 10.39 & 10.44 & 0.53 & \rev{22.3} & 11.09 & 10.38 & 0.16 & \rev{10.8}\\
m12r
 & 9.43 & 9.65 & 0.63 & \rev{4.06} & 9.66 & 9.45 & 0.38 & \rev{0.12} & 10.26 & 9.97 & 0.34 & \rev{2.2} \\
m12w
 & 9.25 & 10.13 & 0.88 & \rev{2.28} &  9.97 & 10.13 & 0.59 & \rev{3.41} & 10.79 & 9.84 & 0.10 & \rev{7.1} \\
\hline
\multicolumn{13}{l}{Notes: Quantities measured within a 30 kpc cubic volume. }\\
\multicolumn{13}{l}{{\textdagger}Star formation rates averaged over a \rev{10} Myr period, ignoring stellar evolutionary mass loss.}
    \end{tabular}
\end{table*}
 
The contribution of the FRB DM from the IGM vs. host galaxy is uncertain, but most works-to-date suggest the host galaxy DM is subdominant to the IGM contribution \citep{2021ApJ...906...95Z, 2020Natur.581..391M}. The DM distribution likely takes on a lognormal form in the ISM, which has also been observed in ionized and neutral gas tracers in the local MW ISM \citep{2008ApJ...686..363H,bialy2020driving,bialy2019chemical,Imara_2016,Herron_2016}.

Studies using cosmological simulations 
\citep{2020ApJ...900..170Z,Theis2024} estimate the halo/ISM DM  of the host galaxies to be $\sim 30-100$ pc cm$^{-3}$ at $z=0$, and increases with increasing redshift. 
Recently, \citet{2023MNRAS.518..539M} used Illustris and IllustrisTNG to estimate 
 the dispersion measure of millions of mock FRB events and study the contribution by the host galaxy and parent halo, $\rm{DM_{host}}$, between redshift $z=0$ and $z=2$. 
 They studied mocks events with two populations: one that traced the SFR, \ie
 ``a young progenitor model'', and the other was assumed to trace the stellar mass, \ie associated with ``old progenitors.'' 
 
  \citet{2023MNRAS.518..539M} found differences between old and young progenitor populations as well as trends with redshift and stellar mass of galaxies, and the median values of $\rm{DM_{host}}$ with the FRB population tracing the star formation rate (SFR) were 179 and 53 $\rm{pc \,cm^{-3}}$ for galaxies in the TNG100 and Illustris simulations, respectively. For the population tracing stellar mass, they found smaller values of DM:  63 and 31  $\rm{pc \,cm^{-3}}$ for TNG and Illustris.  Using these findings, \cite{2023arXiv230316775L} investigated the FRB energy function with assumptions of DM host variations, comparing a constant host DM model of 50 $\rm{pc \,cm^{-3}}$ to a lognormally distributed DM model constrained with DM values from cosmological simulations \citep{2023MNRAS.518..539M,2020ApJ...900..170Z}. In general, because the observed FRB total DM$_{\rm obs}$ are much higher than this, such studies using cosmological simulations suggest that \textit{on-average} most of the DM contribution from FRBs should come from the IGM.  However, their work shows that uncertainty in host-galaxy DMs makes it difficult to make strong conclusions relating to progenitor models (\eg do FRB occurrence rates follow star formation rate density or stellar mass density).

However, there is some evidence that FRBs have significantly larger host-galaxy DMs than suggested by cosmological simulations.
For example, repeating FRB 20190520B has an estimated host-galaxy ${\rm DM_{host}} =430^{+140}_{-220} $ pc cm$^{-3}$ (after accounting for foreground galaxy clusters) and a redshift of $z=0.24$ \citep{2022Natur.606..873N,Koch_Ocker_2022,2022arXiv220211112A, 2023ApJ...954L...7L}, which provides a cautionary example of interpreting high-DM FRBs as having high redshifts.

While studies employing cosmological simulations have provided insight into the DMs of a large range of potential host halos that trace different larger-scale environmental conditions and redshifts, they may not provide a detailed picture of variations on smaller ISM scales. The aforementioned results may be significantly biased by poor resolution of the ISM and the particulars of the sub-grid modeling of the ISM, \eg the \cite{Springel2003} effective equation of state.  The distribution of interstellar and halo gas and electrons that would alter DM are strongly affected by the choices of star formation prescription, gas cooling, cosmic rays, and feedback from massive stars and active black holes \citep{Tasker2008, Hummels2013, Ji2020, Su2020, Ponnada2022}. Thus, it behooves us to use the highest-resolution cosmological simulations to forward model DM predictions, owing to the small-scale perturbations to the free electron content in the regions local to possible FRB progenitors. 

This paper uses the higher-resolution FIRE-2 simulations to investigate the DM across simulated host galaxies.  Similar to \cite{2023MNRAS.518..539M}, we focus on two populations of progenitors: a young population that resides in HII regions and an older population outside of HII regions.  FIRE-2 galaxies have an advantage over large box cosmological simulations galaxy populations' in that they directly resolve HII regions and the ambient ionized ISM down to pc-scales, while at the same time evolve in a self-consistent cosmological context. The results of these analyses indeed indicate an increase in the predicted DM from the host galaxy, relative to the coarser resolution TNG simulations presented in \cite{2023MNRAS.518..539M}.

This letter is organized as follows: in Section \ref{sec:meth} we describe the FIRE-2\footnote{\url{https://fire.northwestern.edu/}} simulations and our method for calculating dispersion measures from the stellar populations in the galaxy snapshots, agnostic to any FRB progenitor model.  In Section \ref{sec:analysis}, we describe our results, comparing with the observed set of FRB DMs, and exploring the possible effects of inclination, followed by a discussion in Section \ref{sec:disc}. Lastly, we summarize and conclude in Section \ref{sec:sum}.

\section{Simulations, Observations, \& Methods}\label{sec:meth}

\begin{figure*}
    \centering
    \includegraphics[width=1\linewidth]{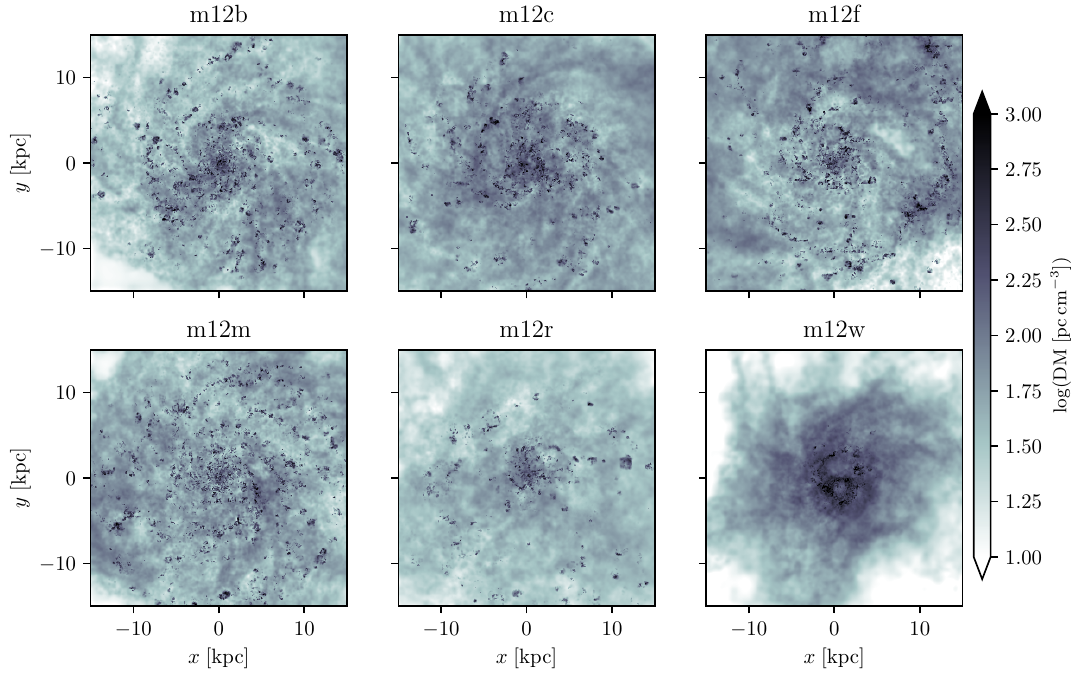}
    \caption{Spatially resolved (100 pc smoothing) face-on maps of dispersion measure (DM) in the six FIRE-2 simulated galaxies analyzed in this paper, at $z=0$. Free-electron column is projected 30 kpc through the analysis volume ($\pm$15 kpc from the galactic midplane). H{\scriptsize II} regions surrounding young massive stars are clearly visible throughout the disks of all six galaxies.  \textbf{m12r}, being $\sim$0.5-0.8 dex less massive in stellar mass, hosts visibly fewer star-forming sites.  \textbf{m12w} has a significantly more compact gas disk than the other galaxies, contributing to relatively high DMs compared to the other halos. Images at $z=1$ \& 2 are found in Appendix~\ref{sec:append:DMmaps}.}
    \label{fig:z0maps}
\end{figure*}

We use simulations from the FIRE-2 public data release \citep{Wetzel2022}, specifically the $z =$ 0, 1 \& 2 snapshots of six halos that form Milky Way mass galaxies by late times (dark-matter halo masses of $M_{200m} = 1 - 2 \times 10^{12}$ M$_\odot$ at z = 0).  Table~\ref{table:gal_props} highlights general physical properties of these galaxies at the redshifts analyzed in this letter. The FIRE-2 cosmological zoom-in simulations of galaxy formation are part of the Feedback In Realistic Environments (FIRE) project, ran using GIZMO \citep{Hopkins2015:gizmo} in its mesh-free Lagrangian Godunov (meshless finite mass, MFM) mode, with the FIRE-2 physics model \citep[][details of the model therein]{Hopkins2018:fire}.  The mass resolution (\ie minimum baryonic element/particle mass) of these simulations is 7100 M$_\odot$, with a minimum (adaptive) force softening length of $\lesssim 1$ pc.  A standard flat $\Lambda$CDM cosmology is assumed, with $h\approx 0.7$, $\Omega_{\rm M} = 1 - \Omega_\Lambda \approx 0.27$, and $\Omega_{\rm b} \approx 0.046$. We reiterate here for the reader some details of the star formation and feedback prescriptions germane to gas ionization in the simulations:

Stars are formed in the simulations when gas is dense ($> 10^3$ cm$^{-3}$), molecular ($f_{\rm H_2} > 0.5$, per the \citealt{Krumholz2011} empirical fit for molecular gas fraction as a function of column density and metallicity), locally self-gravitating (virial parameter $\alpha_{\rm vir} < 1$), and Jeans-unstable below the resolution scale.  The star particles spawned from the gas elements are treated as single stellar populations (SSPs) with a known age, metallicity and mass.  Stellar feedback quantities are derived from the STARBURST99 \citep{Leitherer1999} model, assuming a \citet{Kroupa2002} IMF.

The ionization state of the gas is self-consistently tracked, following heating and cooling from: supernovae (Type Ia and II), stellar mass loss (OB/AGB winds), photoionization, photoelectric heating, hydrodynamic terms, metal lines, dust collisions, free-free emission, Compton cooling/heating, gas collisional excitation/ionization, recombination, and fine-structure and molecular lines. 

\subsection{Calculating Dispersion Measures from the FIRE-2 Simulations}


We calculate the dispersion measure to each star particle in the snapshot by a simple raytracing scheme.  We first orient the galaxy face-on (we consider the effects of inclination in \S~\ref{sec:inclination}) using the stellar angular momentum (as in \citealt{Orr2020} and \citealt{Orr2023}).\footnote{At $z=0$, all six galaxies have well-defined disks and are clearly rotation-supported.  However, for the $z=1$ \& 2 snapshots, ``face-on'' is poorly defined as the galaxies are all disordered, irregular, dispersion-supported objects.}  We then trace rays originating from each star particle out of the galaxy along this defined z-axis, summing the free electron column of the ionized gas elements along the line of sight (\ie Eq.~\ref{eqn:dm}).  To focus our analysis on the host-galaxy contribution to the DMs, we sum only gas elements within $\pm 20$~kpc of the galactic midplane.  Thus, this is a lower limit to the DM contributions of the host galaxy, as its circumgalactic medium will add some minor amount of DM unaccounted for here. We generally do not add a $1/(1+z)$ cosmological redshift factor to our DMs, and present the ``intrinsic'' host-galaxy DMs.

\subsection{Observed FRB Dispersion Measures}\label{sec:obs}
To compare our calculated host-galaxy DMs against the observed distribution of FRB DMs, we searched for reported FRB events: as of \rev{July 31, 2024}, 43 FRBs with measured DMs (${\rm DM_{Obs}}$), localized redshifts \rev{with host galaxy identifications}, and estimates for Milky Way foreground DM  (${\rm DM_{MW,ISM} + DM_{MW,Halo}}$) were found \rev{in the literature}, which we took as a representative population. 
\rev{Appendix~\ref{sec:append:frbobs} lists the details of this FRB sample, including details of the host galaxies and localization references. As well, the redshift and dispersion measure (minus MW contributions) distributions can be found therein.}

\rev{For consistency in our comparison, we have taken the reported `full' DM values (\ie that of Eq.~\ref{eqn:DM_obs}) as reported in the literature for each of the FRBs.} 
We subtract \rev{the} estimated Milky Way contribution as provided \rev{by the source (generally this has been from the {\scriptsize NE2001} model of \citealt{Cordes2002})}, \ie ${\rm DM_{Obs} - DM_{MW,ISM} - DM_{MW,Halo}}$. These FRBs have been localized to galaxies with measured redshifts too, and so we are able to subtract an estimated mean IGM contribution following \citet{Yang_2017}:
\begin{equation}\label{eqn:IGM}
    \left< {\rm DM_{IGM}} \right> = \frac{3cH_0 \Omega_b f_{\rm IGM}f_e}{8 \pi G m_p} \int^z_0 \frac{1+z'}{\sqrt{\Omega_m(1+z')^3+\Omega_\Lambda}} dz' \; ,
\end{equation}
where we have adopted the standard \citet{2016A&A...594A..13P} cosmology ($H_0=67.7$ km/s, $\Omega_b = 0.049$, $\Omega_m = 0.31$), $f_e = 7/8$ (see \citealt{Yang_2017}, where we and they both assume full hydrogen and helium ionization in the IGM below $z \sim 3$), $f_{\rm IGM} = 0.83$ \citep{Shull2012}, and $c$, $G$ and $m_p$ are the speed of light, gravitational constant, and proton mass, respectively. The resulting inferred observed host-galaxy FRB DMs are thus of the form:  DM$_{\rm Host} = (\rm{DM_{obs} - DM_{MW,ISM+Halo} - \left< DM_{IGM}\right> )\cdot (1+z)}$. 
Generally, we represent these observations in our Figures with a shaded band covering the 5-95\% data range for the observed DMs, and show the median value (\rev{55.1, 545, \& 157.6} pc~cm$^{-3}$, respectively). \rev{Except for one observed FRB,} we do not break out the observed values by redshift, and compare them against all simulated redshifts as a means to discuss the general constraints of host-galaxy DMs at present.  \rev{The only exception being FRB 20220610A reported by \citet{Gordon2024}, having been localized to a star-forming galaxy at $z \approx 1.017$, which we compare directly against the FIRE-2 simulations at $z =1$.  For the purposes of comparing this FRB, we plot their predicted $\rm DM_{Host}/(1+z) = 515^{+122}_{-272}$ pc cm$^{-3}$ directly (not recalculating their reported mean IGM contribution, and the resulting shift in DM$_{\rm Host}$).}

\section{Analysis}\label{sec:analysis}

\begin{table*}\caption{Dispersion Measure Percentiles at Integer Redshift $z = 2-0$ of All Star Particles in FIRE-2 Galaxies}\label{table:DMs}
\centering
\begin{tabular}{cccccc}
\hline
\multicolumn{1}{l}{}   & \multicolumn{1}{c}{Ages} & 5\%  & 50\% & Mean & 95\% \\ \hline
\multirow{4}{*}{$z=2$} & \multirow{2}{*}{Young (\textless{}10 Myr)}  & 92.9 & 285  & 525  & 1782 \\
& & (31.0) & (95) & (175) & (594) \\
& \multirow{2}{*}{Old (\textgreater{}10 Myr)} & 31.6 & 157  & 205  & 504  \\ 
& & (10.5) & (52.3) & (68.3) & (168) \\ \hline
\multirow{4}{*}{$z=1$} & \multirow{2}{*}{Young} & 56.0 & 266  & 795  & 3420 \\
& & (28.0) & (133) & (398) & (1710) \\
& \multirow{2}{*}{Old} & 23.4 & 79.7 & 107  & 240  \\ 
& & (11.7) & (39.9) & (53.5) & (120) \\ \hline
\multirow{4}{*}{$z=0$} & \multicolumn{1}{c}{Young (face-on)}                                                                & 21.6 & 93.6 & 210  & 787  \\
                       & \multicolumn{1}{c}{Old (face-on)}                                                                  & 8.4  & 43.0 & 73.8 & 230 \\
                        & Young (orientation avg.d)& 30.1 & 157 & 328  & 1203  \\
                       & Old (orientation avg.d) & 11.5  & 84.4 & 149 & 488    \\ \hline
\multicolumn{2}{c}{Host galaxy-localized FRBs} &\multirow{2}{*}{\rev{55.1}} & \multirow{2}{*}{\rev{157.6}} & \multirow{2}{*}{\rev{193.8}} & \multirow{2}{*}{\rev{545.0}} \\
\multicolumn{2}{l}{$(1+z)$(DM$_{\rm Obs}$-DM$_{\rm MW}$-$\left<{\rm DM}_{\rm IGM}\right>$)}   &&&& \\ \hline

\end{tabular}\tablecomments{DMs all in units of pc cm$^{-3}$, all values in parentheses include 1/(1+$z$) cosmological redshift term.}
\end{table*}

\begin{figure}
    \centering
    \includegraphics[width=1\linewidth]{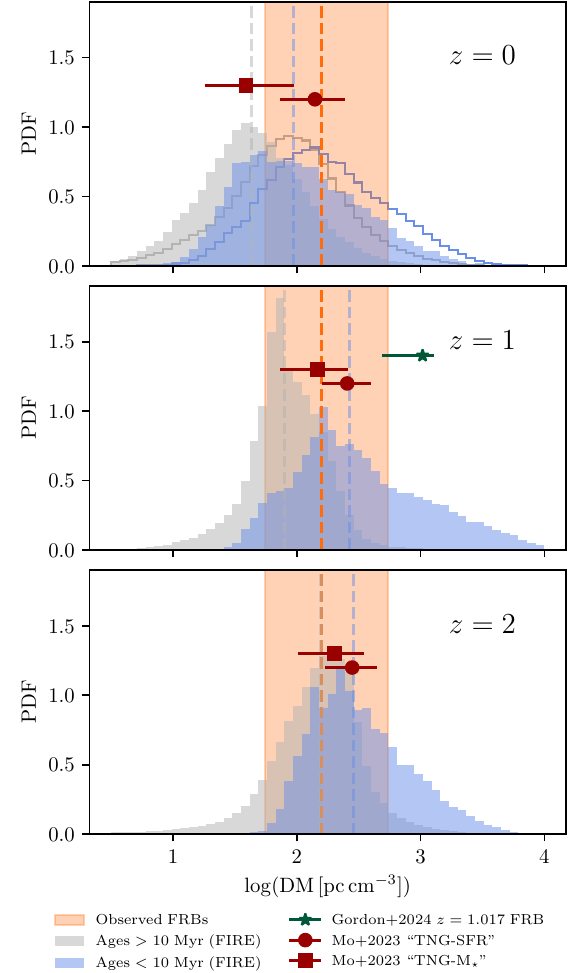}
    \caption{Distribution of intrinsic host-galaxy dispersion measure for all stars in the six FIRE-2 simulations analyzed here at $z = 0,1,$ \& 2.  Blue (grey) shaded histogram shows DM distribution for stars younger (older) than 10 Myr, with vertical dashed blue (grey) line denoting median value.  Unfilled blue (grey) histograms represent orientation-averaged (filled is `face-on') DM distribution for the stellar population(s) in the disks at $z=0$, median value not shown here.  Orange shaded band and vertical dashed line show the 5-95\% interval and median value, respectively, for observed inferred FRB population host-galaxy DMs (with estimates for MW ISM+halo and mean IGM contributions subtracted; see \S~\ref{sec:obs} for data sources). Cardinal points and errorbars show predicted host-galaxy DMs from IllustrisTNG simulations \rev{\citep[][]{2023MNRAS.518..539M}}, with circle \& square points denoting range of \rev{star-forming host galaxy DMs (``TNG-SFR'', with specific SFRs ($\dot M_\star/M_\star) > 10^{-11}$ yr$^{-1}$) and stellar mass-weighted host galaxy DMs (``TNG-M$_\star$'', weighted by stellar mass in their sample)} sightlines, respectively. At all redshifts, young populations are embedded in much higher ionized gas columns (\ie DMs) by $\sim$0.3-0.5~dex, compared to older stars in the simulations. For $z=0$ disk galaxies, inclination effects can significantly increase (by a factor of $\sim$2) DMs, especially compared to other simulations.
    } 
    \label{fig:stackedDMPDFs}
\end{figure}

\begin{figure*}
	\includegraphics[width=\textwidth]{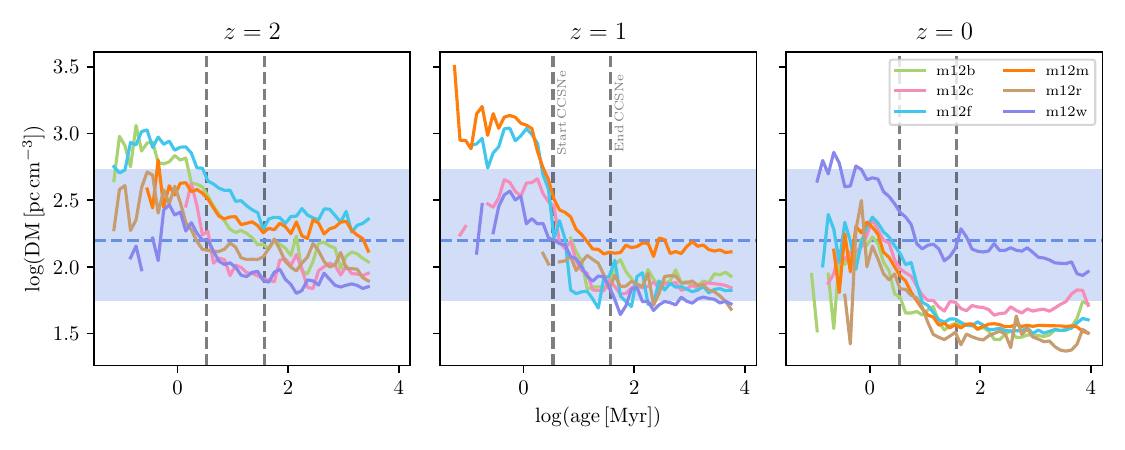}
    \caption{Median 
    free electron column density (intrinsic dispersion measure, DM) from FRB sightlines towards stars as a function of stellar age in six Milky Way mass FIRE-2 simulated galaxies at $z=2-0$. At $z=0$, all six galaxies have disk morphologies and are oriented face-on.  Vertical dashed lines show beginning (3.4 Myr) and end (37.5 Myr) of core-collapse SNe in FIRE-2 model. Horizontal blue shaded band, and dashed line, show 5-95\% range and median value of observed inferred FRB host-galaxy DM distribution (see \S~\ref{sec:obs} for data sources). Highly bursty star formation in all six galaxies at $z = 1$ \& 2 produce significant variance in their median DMs as a function of age (in some cases temporarily suppressed star formation creates gaps).  Similar evolution and normalization appears across all six galaxies at all three redshifts.  Very young stars (ages $<$ 1 Myr) have the highest DMs, and DMs rapidly fall to a plateau after $\sim$40 Myr.} 
    \label{fig:agesNscatters}
\end{figure*}

\begin{figure*}
	\includegraphics[width=\textwidth]{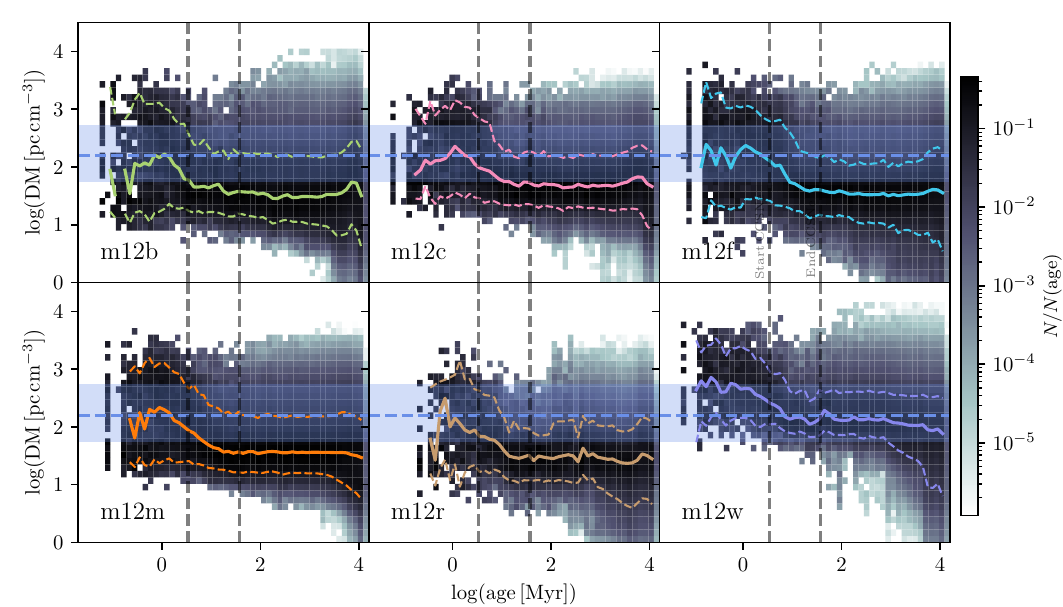}
    \caption{Distributions of dispersion measure (DM, \ie free electron column density) in six Milky Way mass FIRE-2 simulated galaxies at $z=0$, their disks oriented face-on.  Colored lines show median value of DM as a function of star particle age (dashed lines show 5-95\% data range), as right top panel of Fig.~\ref{fig:agesNscatters}. Colormap shows distribution of DMs in age bins, weighted such that the the PDF at each age bin is shown. Horizontal blue shaded band, and dashed line, show 5-95\% range and median value of observed inferred FRB host-galaxy DM distribution (see \S~\ref{sec:obs} for data sources). Vertical dashed lines show beginning (3.4 Myr) and end (37.5 Myr) of core-collapse SNe in FIRE-2 model.  Clear, similar evolution is seen in all six galaxies, with youngest (age $<$ Myr) stars most deeply embedded in ionized gas regions. By the time CCSN feedback ends, stars are under a similar electron column (though with a large scatter).  DM host-galaxy contribution for young stars (ages $<$ 10 Myr) can generally be sufficient to explain observed FRB DMs.  And, though less-likely, DMs from older stars have full overlap with the observed FRB DM range.}
    \label{fig:z0dists}
\end{figure*}

\begin{figure*}
	\includegraphics[width=\textwidth]{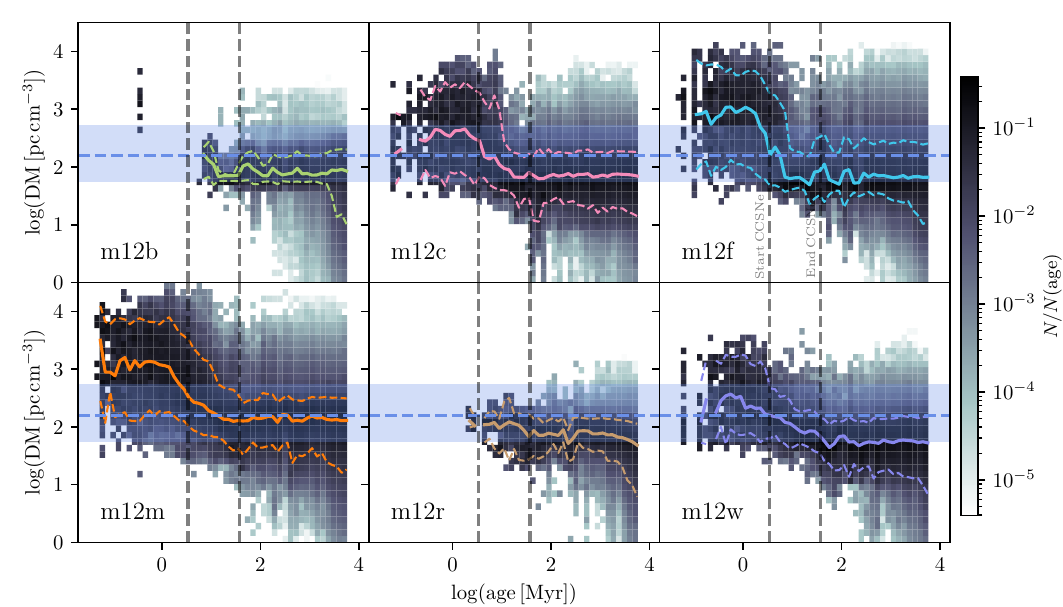}
    \caption{Similar to Figure ~\ref{fig:z0dists} at $z=1$. None of these galaxies have disk morphologies at this redshift. Nearly the same trend is seen as at $z=0$, however the distributions of young stellar ages are incomplete owing to highly bursty star formation in these galaxies. This is seen especially strongly in \textbf{m12b} and \textbf{m12r}, which have just gone through a star burst and have (nearly) fully suppressed star formation for the last 3$-$5 Myr.}
    \label{fig:z1dists}
\end{figure*}

\begin{figure*}
	\includegraphics[width=\textwidth]{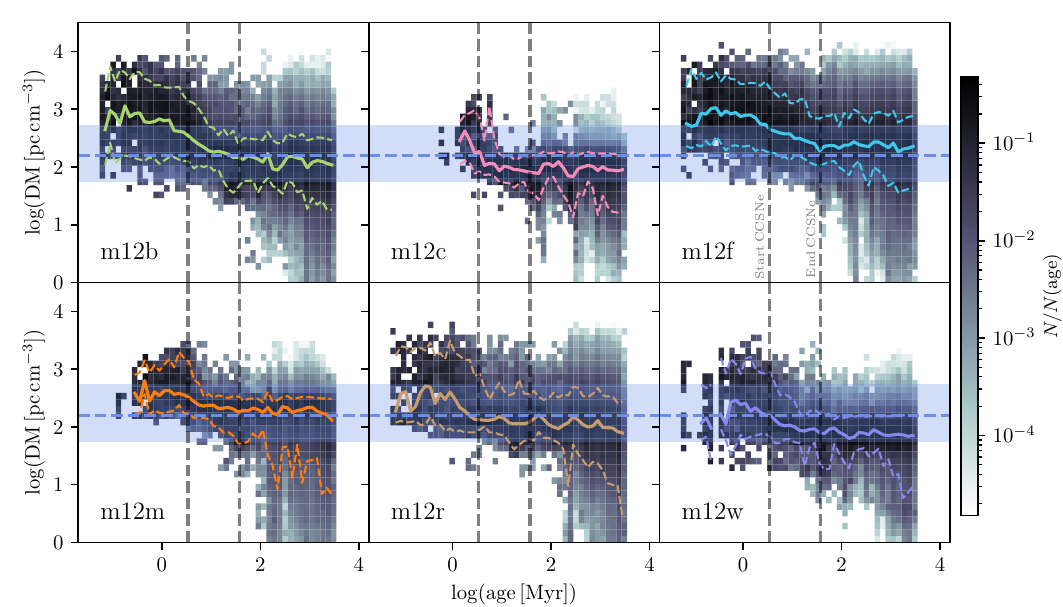}
    \caption{Similar to Figure ~\ref{fig:z0dists} at $z=2$. None of these galaxies have disk morphologies at this redshift, and the results are almost indistinguishable from $z=1$.  Highly bursty star formation at this time produces a dearth of young stars (ages $<$ $\sim$Myr) in \textbf{m12c}.}
    \label{fig:z2dists}
\end{figure*}


Figure~\ref{fig:z0maps} shows maps of the free-electron column (\ie dispersion measure, DM) on 100~pc scales in the six simulated galaxies analyzed here at $z=0$, where we have summed the entire free-electron column 15~kpc\footnote{We did not sum the full $\pm 20$~kpc analysis volume purely for visualization reasons.  Quantitatively, there is little to no difference as the column of ionized gas with $15 < |z {\rm [kpc]}| < 20$ is very low relative to the contribution from the innermost CGM ($2 \lesssim |z {\rm [kpc]}|< 15$) and full ISM ($|z| \lesssim 2$ kpc).} above and below the disk mid-plane in each simulation (additional figures in Appendix~\ref{sec:append:DMmaps} show DM maps of these simulated galaxies at $z=1$ and $z=2$ at the same spatial resolution).  H{\scriptsize II} regions, easily visible as compact high-DM areas, are distributed across the galaxy disks and are indicative in their number of the general level of star formation in each simulation: \textbf{m12r}, being less massive, has generally lower levels of star formation and fewer H{\scriptsize II} regions; and \textbf{m12w} has a dramatically more compact gas disk, which is misaligned relative to the angular momentum of the stellar disk due to a recent LMC-mass merger, resulting in high overall DMs across its face.  All of the disks broadly host full-column DMs on the order of 100-1000 pc cm$^{-3}$ in their star-forming disks.

Focusing on the DMs measured to each star particle (ray-tracing from the particle to a height 20 kpc above the galaxy mid-plane), rather than the full free-electron column through the galaxy, we generate PDFs for the DM distribution towards all the star particles in all six galaxies at $z=0,1$ \& 2 broken out for young (ages $< 10$ Myr) and old (ages $> 10$ Myr) populations, shown in Figure~\ref{fig:stackedDMPDFs}.  
In Appendix~\ref{sec:append:DMmaps}, we present the DM distribution in each individual galaxy at $z=0,1$ \& 2. We compare the DM distributions from the FIRE galaxies with the observed inferred FRB host-galaxy DMs (the total DMs less the MW-foreground and mean IGM estimates).  Unsurprisingly, young stars (ages $< 10$ Myr) in the simulations produce hot, ionized H{\scriptsize II} regions around themselves which cause their DM distribution to be shifted to higher values.  Broadly, there is significant overlap between the FIRE-2 host-galaxy DMs at all redshift and the observed FRB DM range (more so with the young-star DM distributions). Given that the observed DMs are presumed to be just DM$_{\rm Host}$, {significant overlap of FIRE DM$_{\rm Host}$ distributions allow us to infer possible FRB progenitor objects}.  We also include the outline of the orientation-averaged host-galaxy DM distribution for the $z=0$ snapshots, as those are the only ones that host disks (and thus are sensitive to inclination effects).  Inclination on-average increases the host-galaxy DM distributions by up to a factor of $\sim$2, the effects are explored in more detail in \S~\ref{sec:inclination}.

And with distributions with generally higher host-DMs than previously appreciated, we demand that a larger fraction of the observed DM in unlocalized sources is DM$_{\rm Host}$ instead of DM$_{\rm IGM}$-- the observed unlocalized FRB population may be significantly closer (in redshift/cosmological distance) than typically assumed. There may also be additional non-negligible DM contributions from the localized electrons around the FRB progenitors in the form of a nebula. However, given the uncertainty in what FRB progenitors are, along with the significant shift in overall predicted DMs between young and old stellar populations, our analysis is still plausibly consistent with the various FRBs originating from host galaxies at significant redshift.  Moreover, the evolution in the FIRE DMs between $z= 0,1,$ \& 2 may well be attributed to differences predominantly in galaxy morphology and star formation rates, rather than intrinsic, redshift-dependent, differences.  And so, local galaxies with irregular morphologies or high SFRs/gas fractions (\eg the DYNAMO survey galaxy sample; \citealt{Lenkic2021}) should host similar high-DM$_{\rm Host}$ distributions, potentially accounting for much of the inferred extragalactic contribution to observed unlocalized FRB DMs.

Comparing to host-galaxy DM predictions from IllustrisTNG by \citet{2023MNRAS.518..539M}, we see that the multiphase ISM in the FIRE-2 simulations hosts a far broader range of host-galaxy DMs than the much smoother effective equation of state ISM in the cosmological simulations. \rev{We have subtracted their reported CGM contributions to DM$_{\rm host}$, of 17\% and 40\% for their TNG-SFR and TNG-M$_\star$, respectively.}  The offset in FIRE to higher median DMs relative to their  \rev{star-forming host galaxy DM (``TNG-SFR'', with specific SFRs ($\dot M_\star/M_\star) > 10^{-11}$ yr$^{-1}$) and stellar mass-weighted host galaxy DM (``TNG-M$_\star$'', weighted by stellar mass in their sample)} lines of sight, and the much broader scatter to high (and low) DMs in the wings of the distribution, near $z\approx 0$ and $z = 1$ demonstrates the importance of modeling a highly dynamic multiphase ISM in understanding the environments in which possible FRB progenitors live and evolve.

Table~\ref{table:DMs} lays out the broad properties of the DM$_{\rm Host}$ results from the FIRE simulations at $z = 0,1$ \& 2, and compares against the range of observed FRB DMs.  We list 5\%, 50\% (median), mean (weighted by stellar mass in the simulations) and 95\% DM$_{\rm Host}$ values, broken out for young (ages $<10$ Myr) and old (ages $>10$ Myr) stars in the simulations. For our $z=0$ snapshots, we also average over orientation angle to present results consistent with a population of randomly observed disks (see \S~\ref{sec:inclination}). For the observed FRB sample (sources listed in \S~\ref{sec:obs}), we take the percentiles and mean by equally weighting each observation.

Figure~\ref{fig:agesNscatters} shows both the median 
host-galaxy dispersion measure 
as a function of age for the star particles in the FIRE-2 galaxies at $z =$~0, 1, \& 2 in `face-on' orientations 
(the effects of inclination on DMs in the FIRE disks are explored in \S~\ref{sec:inclination}, and we present the full stacked distribution of DM--stellar ages for each redshift in Appendix~\ref{sec:append:stacks}).  We directly compare the FIRE-2 host-galaxy dispersion measures with the observed inferred FRB population host-galaxy dispersion measure distribution (showing the median and 5-95\% range).  We add vertical lines in age at 3.4 and 37.4 Myr to denote the range of possible stellar lifetimes for core-collapse SN progenitors (specifically, the age range used in the FIRE-2 core-collapse feedback model).  Across redshift, the FIRE-2 Milky Way progenitors show qualitatively similar distributions in dispersion measure as a function of age, despite growing considerably in mass and undergoing dramatic changes to their morphologies (\ie settling into disks between $z \approx 1$ and $z =0$ in this sample).  Though stellar populations older than $\sim$10 Myr have median DMs roughly a 0.5-1 dex lower than the observed inferred FRB host-galaxy DMs, young populations (with ages $\lesssim 10$~Myr) have significantly higher median DMs, owing to their tendency to still be embedded in H{\scriptsize II} regions.  Generally, the DMs begin falling before the onset of core-collapse SN feedback in FIRE, showing that pre-SN feedback (\ie stellar jets, winds, and radiation) more than capable of dispersing the ionized gas regions surrounding young massive stars \citep[see, \eg][]{Grudic2018, Lancaster2021, Grudic2022}. 

For the $z = 1$ and $z = 2$ FIRE-2 snapshots, the young populations have more than enough free electron columns to \emph{completely} account for the observed inferred FRB host-galaxy DMs, and then some. 
That is to say, young populations from lower-mass starbursting galaxies with high gas fractions in the simulations contain sufficient ionized gas columns to entirely explain observed inferred FRB host-galaxy DMs.  Should there exist $z = 0$ analogues of these, FRBs originating from young stellar populations would be of little use as probes of cosmological distributions of the ionized intergalactic medium as they necessarily must be significantly closer to us in the Milky Way than previously estimated for non-localized sources.

\rev{In the $z = 1$ and $2$ snapshots, given their irregular morphology and dispersion supported nature, we clearly see in the DM maps (Figs.~\ref{fig:z1maps} \& \ref{fig:z2maps}) that they are dominated by a few large star-forming regions (rather than the distributed star formation seen in the $z=0$ disks).  Consequently, the host DMs measured for the youngest stellar populations correspond directly, or nearly so, to the recent SFR of the entire galaxy (cf.~Table~\ref{table:gal_props}), as the global intensity of the star formation is much more tightly related to the radiation field seen by the gas around the youngest stars.  Whereas, the distributed nature of star formation in the disks at $z=0$ means that the global star formation rate is not as important in setting the local radiation field seen by gas around young stars, and the individual star forming regions are significantly smaller producing smaller ionized gas columns around young stars/star-forming regions.}

Notably, \textbf{m12w}, a heavily disrupted disk galaxy having just underwent a major merger-driven starburst at $z =0$ with a highly compact gas disk, too hosts a stellar population with median host-galaxy dispersion measures sufficient to explain observed FRB DMs at nearly all stellar ages.  This \rev{challenges the} assumption that only the lower-mass, star-bursting, high-gas fraction progenitors can produce such local ISM conditions.

We also explore the scatter in host-galaxy DM as a function of stellar age in Figures~\ref{fig:z0dists} - \ref{fig:z2dists}.  The DMs towards the youngest stars, across all redshift in the FIRE simulations, have the highest amount of scatter (\ie 5-95\% range) at fixed age, highlighting the heterogeneity of local ISM conditions and gas columns in and around star-forming regions.  As the stars age and mix in with the wider galaxy with ages between $\sim$10 Myr and $\sim$1 Gyr, the scatter in their free electron columns falls by roughly a factor of two.  However, the oldest stars (ages $\gtrsim$~Gyr) again exhibit a higher degree of scatter in their DMs, owing to a larger portion of that population existing on isotropized halo orbits or in the dynamically hotter thick disk (in the case of the $z =0$ snapshots).  These two classes of orbits have columns in some cases cutting across the whole of the ISM, or are above the disk and have essentially no ISM (only CGM) in their line of sight.

The DM distributions at $z=0$ (Figure~\ref{fig:z0dists}) show a consistent evolution, with nearly identical tracks as the stars age out of local H{\scriptsize II} regions into the ambient ISM (with the aforementioned exception of \textbf{m12w}, in terms of absolute normalization).  The tail to high DMs in the oldest population can be attributed to chance alignments with H{\scriptsize II} regions after the population is well-mixed into the ISM.  The slight lack of a tail to high DMs in intermediate age (ages $\sim$ 10 Myr) stellar populations points to the anti-correlation of currently SN feedback-producing regions with actively star-forming regions (which produce H{\scriptsize II} regions in short order), and is thus not terribly surprising that they have, on-average, the least scatter in DMs out of all age bins.

The higher redshift snapshots in Figures~\ref{fig:z1dists} and \ref{fig:z2dists} have similar DM distribution characteristics to the $z=0$ snapshots, however the lack of a disk morphology and globally bursty star formation is on display.  Several of the galaxies are ``post starbursts'' (narrowly defined here as having just had a starburst $\sim$Myr ago) and hosting no ages $<$ Myr populations; these include \textbf{m12b} and \textbf{m12r} at $z=1$ and \textbf{m12c} at $z=2$.  Given overall higher specific star formation rates and gas fractions, it is expected that the galaxies produce higher ambient electron densities and larger H{\scriptsize II} complexes, increasing the DMs in all age bins at these early times \citep[cf. observational work by:][]{Shimakawa2015, Reddy2023}.

\subsection{Effects of Inclination}\label{sec:inclination}

\begin{figure*}
	\includegraphics[width=\textwidth]{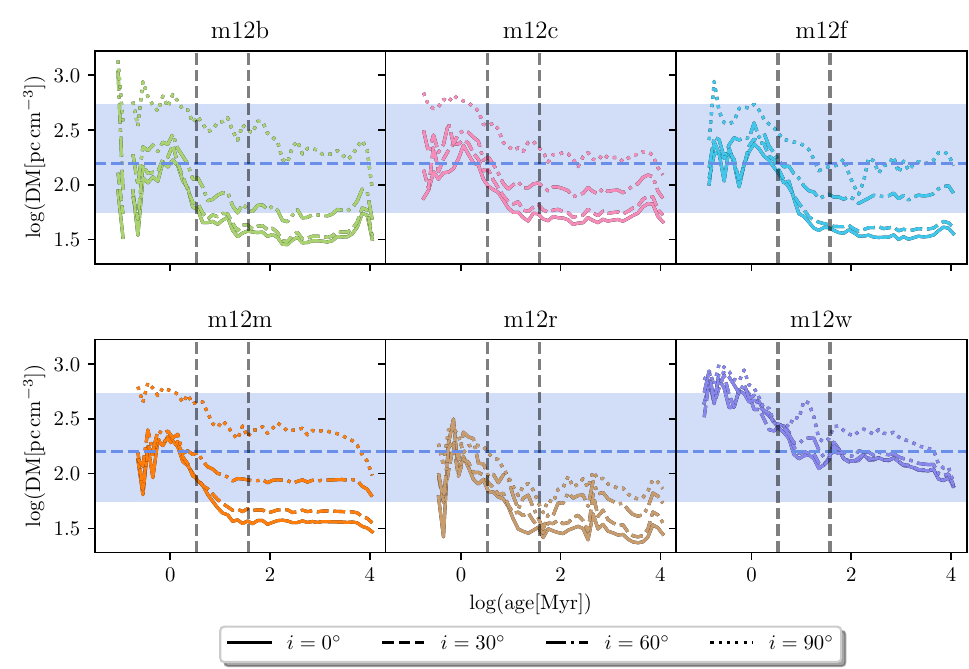}
    \caption{Median dispersion measure (DM, \ie free electron column density) as a function of age in six Milky Way mass FIRE-2 simulated galaxies at $z=0$ at various inclination angles, style as Fig.~\ref{fig:agesNscatters}. \rev{Blue shaded band, and dashed line, show 5-95\% range and median value of observed inferred FRB host-galaxy DM distribution (see \S~\ref{sec:obs} for data sources).} At $z=0$, all six galaxies have disk morphologies. Below inclination angles of $\sim$60$^\circ$, only a small change ($<$0.3~dex) in electron columns are seen, and only for the older stellar populations (ages $\gtrsim$ 10 Myr).  Edge-on (and very highly inclined) systems exhibit significantly flatter DM distributions with age, and overall higher electron columns (by 0.5-1.5~dex across all age bins).}
    \label{fig:inclination}
\end{figure*}

Previous figures have all shown the host-galaxy dispersion measures as calculated with inclinations $i = 0^\circ$, \ie `face-on', which is a necessarily conservative estimator of gas column in disks.  For non-disk galaxies, all of the non-$z =0$ snapshots in our case, this makes little difference to our estimations as the ISM is roughly isotropic on kpc scales.  However, inclination effects for the calculated dispersion measure of disk galaxies can be quite significant, and the average inclination of randomly oriented disks on the sky is large, at 60$^\circ$.

Figure~\ref{fig:inclination} shows the median DM as a function of age for the $z=0$ FIRE snapshots, at inclinations $i =$ 0$^\circ$, 30$^\circ$, 60$^\circ$, and 90$^\circ$, alongside the observed FRB $\rm DM_{\rm host}$ distribution.  Low inclinations ($i=30^\circ$) do not greatly affect our DM estimates in any of our disks.  And except for extreme inclinations ($i \approx 90^\circ$), the dispersion measures of the youngest stellar populations are not significantly affected due to the fact that the primary contribution to their DMs are the young stars' own (compact, roughly isotropic) H{\scriptsize II} regions, and not the wider ISM.  For stellar populations no longer deeply embedded in their own H{\scriptsize II} regions (ages $\gtrsim$10~Myr), the average-expected inclination $i =60^\circ$ can increase the median estimated DM by $\sim 0.25-0.5$ dex.  This increase is directly in-line with the boost one might expect for the increased path-length through a slab geometry (with constant or Gaussian vertical density profiles) of $\log(l(i=60^\circ)/H) = \log(1/\cos(60^\circ))\approx 0.3$~dex. With a very compact gas disk and high intrinsic DMs, \textbf{m12w} interestingly does not exhibit any significant boost in DM with inclination, except modestly when viewed entirely edge-on.

At extreme inclinations in all our disk galaxies (\ie when viewed edge-on), except for \textbf{m12w}, the median DMs across all stellar ages are significantly boosted by $\sim 0.5-1$~dex.  In these cases, nearly all of these FIRE galaxies host stellar populations (across all age bins) with DMs sufficiently high to \emph{completely account and then some} for the median observed inferred FRB host galaxy DMs.  
Given that at least half of spiral galaxies in the nearby universe have high inclinations ($i > 60^\circ$), this would suggest that a non-negligible fraction of FRBs originating from $\sim$$L_\star$ disk galaxies may be highly attenuated/scattered by highly inclined disks, biasing our observational sample to less inclined systems. 
Moreover, such highly inclined systems are not particularly constraining of the FRB source population, given the quite flat distribution of host DMs with age.

\subsubsection{Orientation-averaged Stacked DM-host PDFs}
\begin{figure}
	\includegraphics[width=0.5\textwidth]{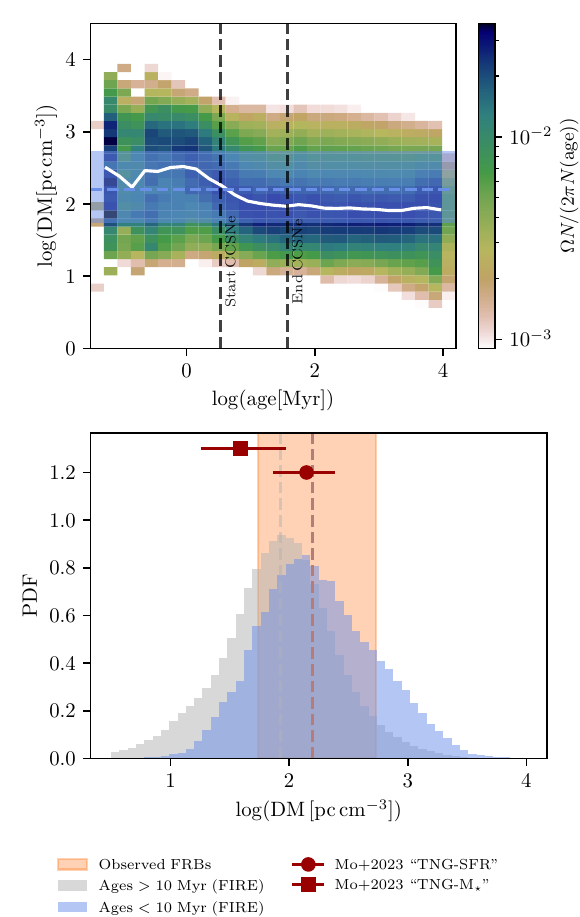}
    \caption{\textit{Top panel:} orientation-averaged dispersion measure (DM, \ie free electron column density) as a function of age in six Milky Way mass FIRE-2 simulated galaxies at $z=0$ stacked together, style as Fig.~\ref{fig:z0dists}. \textit{Bottom panel:} orientation-averaged dispersion measure (DM, \ie free electron column density) PDFs for the six FIRE galaxies at $z=0$ stacked together, in the style of Fig.~\ref{fig:stackedDMPDFs}. Orientation effects reduce the difference between youngest and older stellar populations, as the increased pathlengths through the ambient ionized ISM increase DMs for old stars. Young stars ($\sim$10 Myr old), appear to align best with inferred observed host-galaxy FRB DMs, when orientation-averaged (disk inclination) effects are considered for the simulated FIRE spiral galaxies.}
    \label{fig:inclination_weighted}
\end{figure}

To further explore the effects of inclination, we stack together the different orientations of the FIRE disks at $z=0$, weighting by their relative solid angle on $2 \pi$ (taking the inclinations $i = 0^\circ$, 30$^\circ$, 60$^\circ$, 90$^\circ$ and calculating the area of the bands between them). Figure~\ref{fig:inclination_weighted} shows the result, for both the distribution of host-galaxy DMs as a function of stellar age (white line showing the weighted median), and the PDFs themselves, compared against inferred observed host-galaxy FRB DMs, and those calculated in the IllustrisTNG cosmological simulations by \citet{2023MNRAS.518..539M}.  We again find that the PDFs of host-galaxy DM are wider than previously appreciated, and that younger stars ($\sim$10 Myr old) appear to agree best with the observationally inferred $\rm DM_{\rm host}$
when the effects of orientation (inclination) are averaged over.




\section{Discussion}\label{sec:disc}
In this paper we have explored the relationship between free electron column densities and stellar ages within six Milky Way mass progenitor galaxies, utilizing the FIRE-2 simulations. This exploration reveals consistent DM trends across different redshifts and for galaxies with disk-dominated or irregular morphologies.  Notably, our findings demonstrate that median host-galaxy DMs in FIRE-2 galaxies significantly exceed those predicted by other cosmological simulations, such as Illustris and TNG \citep[][respectively]{Vogelsberger2014, Pillepich2018}, likely due to the fact that such cosmological simulations do not resolve HII regions and use an effective equation of state to model the ISM (\ie the \citealt{Springel2003} model). Such discrepancies highlight the crucial role of modeling the local interstellar medium conditions of the host galaxy in order to constrain FRB  DMs.

The broad distribution (particularly, the `fat tails') of host-galaxy DMs for older stars in the FIRE simulations makes it difficult to definitively argue that young massive stars alone must be FRB progenitors.  However, the identification of a `high floor' in host-galaxy DMs imposes stricter limits on the cosmological distance of detected FRBs and suggests the host-galaxy DM term might play a more critical role than previously thought. These insights critically refine our understanding of potential FRB source populations and the maximum distances of these enigmatic phenomena.

The `high floor' in median host-galaxy DMs of $\sim$100 pc cm$^{-3}$ (see Fig.~\ref{fig:inclination_weighted}) adds more stringent constraints to the maximum redshift/distance of detected FRBs.  Considering a broader reported non-localized FRB sample: as of April 17, 2024, 804 FRBs with measured DMs (${\rm DM_{Obs}}$) and estimates for Milky Way foreground DM  (${\rm DM_{MW,ISM} + DM_{MW,Halo}}$) were found on the server. Originating from TNS FRB Report (FR) 2020 IDs \#891, 892, 928, 934, 1613, 2395, 2470, 2481, 2695, 2752, 3517, 3759,
TNS FRB Report (FR) 2021 IDs \#208, 421, 666, 990, 1346, 2007, 2008, 2276, 3302, 3632,
 TNS FRB Report (FR) 2022 IDs \#11, 28, 444, 693, 1110, 1785, 2269, 3214,
and TNS FRB Report (FR) 2023 IDs \#412, 419, 2364, 2635.  This sample, after subtracting the MW foreground, had a median IGM+host-galaxy dispersion measure of 391 pc~cm$^{-3}$.

Considering the expected mean IGM dispersion measure, as \citet{Yang_2017} \rev{also} do to second order \rev{(Eq.~\ref{eqn:IGM} being the exact integral)}, we find:
\begin{equation}\label{eqn:IGMapprox}
    \left< {\rm DM_{IGM}} \right> \approx \frac{3cH_0 \Omega_b f_{\rm IGM}f_e}{8 \pi G m_p} \left[ z + z^2 \left( \frac{1}{2} - \frac{3\Omega_m}{4}\right)\right] \; ,
\end{equation}
where we have adopted the standard \citet{2016A&A...594A..13P} cosmology ($H_0=67.7$ km/s, $\Omega_b = 0.049$, $\Omega_m = 0.31$), $f_e = 7/8$ (see \citealt{Yang_2017}, where we and they both assume full hydrogen and helium ionization in the IGM below $z \sim 3$), and $f_{\rm IGM} = 0.83$ \citep{Shull2012}. Taking the `typical' (median) DM excess (\ie host-galaxy DM + IGM DM) from the observations of 391 pc cm$^{-3}$, the `floor' of $\sim$100 pc cm$^{-3}/(1+z)$ from the FIRE spirals implies a maximum redshift of $z\approx 0.36$ (considering the remainder of the excess having come from the IGM), whereas if the typical FRB progenitor comes from a system more akin to one of the non-disky snapshots with a mean `young' host-galaxy DM of $\sim$285 pc cm$^{-3}/(1+z)$ then the implied maximum redshift is only $z \approx 0.175$.  For this calculation here we include the $1/(1+z)$ cosmological term for the host-galaxy contribution to the total DM. The general concern being that if the host-galaxy contribution to the observed DM excesses is larger than typically believed by factors of potentially a few, then the ability of FRBs to constrain \textit{cosmological} models might be statistically significantly more limited than previously thought.

 
We note that the  majority of the galaxy snapshots in our analysis are of highly disturbed irregular gas distributions (all of our $z> 0$ snapshots and \textbf{m12w} at $z=0$), many of which are either undergoing or have just endured a starburst event.  These FIRE galaxies generally settle down into steadily star-forming disks between $z \approx 0.7-0.4$ \citep{Gurvich2023}, and so these disturbed conditions do not typically represent the mode of star formation or galaxy evolution in the $L_\star$-mass FIRE simulations at late times. Various observational works characterizing the host-galaxy properties of localized FRBs have identified large HI kinematic asymmetries \citep{Michalowski2021, Glowacki2023}, that in some cases have been associated with merger events.  Disturbed dense gas kinematics have also been associated with some identified hosts \citep{Hsu2023}, as well as clearly merger-driven starbursts \citep{Kaur2022}, suggesting a pervasive (but not necessarily dominant) phenomenon of FRBs occurring in close relation to highly gravitationally unstable star-forming gas reservoirs. This lends credence to our concern that many FRB progenitors occur in the non-disky, starbursting environments found in the $z \gtrsim 1$ FIRE snapshots, implying much larger host-galaxy DMs than previously estimated by theory (in the form of the aforementioned cosmological simulations).







\section{Conclusions}\label{sec:sum}

In this Paper, we investigated the relationship between free electron column densities and stellar ages in six Milky Way mass progenitor galaxies from the FIRE-2 cosmological zoom-in simulation suite at integer redshifts $z = 2-0$.  Doing so, we provide a unique constraint on measured Fast Radio Burst (FRB) dispersion measure (DM) contributions from a host-galaxy ISM, whilst generally being agnostic to FRB progenitor models.
Our primary findings are as follows:
\begin{itemize}
    \item Across redshift and morphology, nearly identical trends in dispersion measure (free electron column) appear in the FIRE-2 Milky Way mass progenitors.  Initially, stars are most deeply embedded in ionized gas columns with $\log({\rm DM \, [pc \, cm^{-3}]}) \approx $ 1.5-3.5, and, as they evolve, their DMs quickly fall until roughly the end of core-collapse supernova feedback, whereupon they have on-average lower galactic DMs ($\log({\rm DM \, [pc \, cm^{-3}]}) \approx $ 2) for the remainder of their evolution.
    
    \item The median DMs we find in FIRE-2 galaxies are $\sim$2-60 tim larger than the typical DMs through the effective equation of state ISMs of $\approx 50$ pc cm$^{-3}$ found in the Illustris cosmological simulations and $\sim$1-2 times larger than those found in the IllustrisTNG cosmological simulations \citep{2020ApJ...900..170Z, 2023MNRAS.518..539M}, especially when considering inclination effects in disk galaxies.
    
    
    
    \item The `high floor' in median host-galaxy DMs of $\sim$100 pc cm$^{-3}$ for stars in the ``probable progenitor'' stellar age range of $0-100$~Myr adds more stringent constraints to the maximum redshift/distance of detected FRBs. For much older stars (ages $>$100 Myr), the tails of the host galaxy DM distribution can be quite extreme, dependent on orientation, with chance alignments with ionized HII regions, or channels nearly free of ionized gas.
    
    \item Contrary to previous cosmological simulation results, the host galaxy DM in detected FRBs may dominate or contribute equally to the observed DM as the intervening IGM. This statement is softened somewhat by cosmological expansion, the $1/(1+z)$ term, however, disordered/high gas fraction star-forming galaxies (\ie high-$z$ analogues) in the local universe stand to host significantly higher host-galaxy DMs than heretofore appreciated.
\end{itemize}

Taken together, these findings suggest that the distribution of measured FRB host-galaxy DMs might well significantly narrow the possible stellar source population, \ie to moderately young (ages $\sim$3-40 Myr) vs. significantly older (ages $\gtrsim$ 100 Myr) populations.  Moreover, the relatively high $\rm DM_{\rm host}$ ($\sim$100 pc cm$^{-3}$) of older stars in FIRE-2 put a stronger constraint on the upper limit of the cosmological distance of detected FRBs, implying that non-localized FRBs may be cosmologically closer than they have generally been inferred for sources without confirmed redshift.

\begin{acknowledgments}
MEO would like to thank Laure Ellis for fruitful conversations that contributed to interpretations in the paper.  \rev{The authors would like to thank the anonymous referee for comments that greatly helped to improve the manuscript.}
The Flatiron Institute is supported by the Simons Foundation.
BB is grateful for generous support by the David and Lucile Packard Foundation and Alfred P. Sloan Foundation.  CBH is supported by NASA grants 80NSSC23K1515, HST-AR-15800, HST-AR-16633, and HST-GO-16703. This work was performed in part at the Aspen Center for Physics, which is supported by National Science Foundation grant PHY-2210452.
This research has made use of NASA's Astrophysics Data System.
\end{acknowledgments}




\bibliography{dm, library, cca_lib}{}
\bibliographystyle{aasjournal}





\appendix

\section{Compiled host galaxy-localized FRB Observations}\label{sec:append:frbobs}

\begin{figure}
    \centering
    \includegraphics[width=0.49\linewidth]{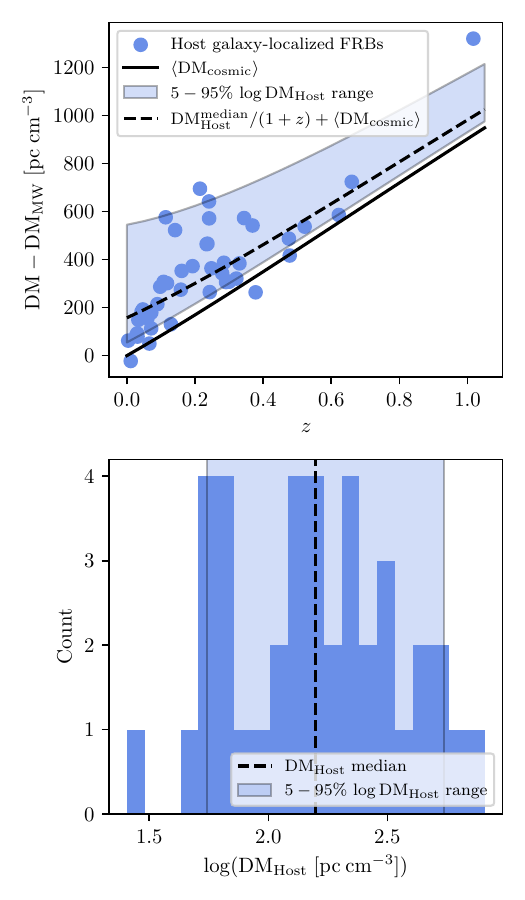}
    \caption{\textit{Top panel:} extragalactic DM excess--redshift relation (`Macquart relation') for observed host-galaxy localized FRBs from Table~\ref{table:frbs}.  Mean IGM DM contribution (solid black line, Eq.~\ref{eqn:IGM}) is plotted, as well as the mean IGM+median host galaxy DM (dashed black line) and mean IGM+5-95\% host galaxy DM range (shaded blue band). \textit{Bottom panel:} inferred host galaxy DM distrbution, $\rm DM - DM_{MW} - \left< DM_{IGM} \right>$, showing median (dashed black line) and 5-95\% range (shaded blue band) of observed host-galaxy localized FRBs. }
    \label{fig:hostgal_dist}
\end{figure}

\begin{deluxetable}{cccccccc}




\tablecaption{Compilation of host galaxy-localized FRBs used for comparison against FIRE-2 simulations}


\tablehead{\colhead{FRB ID} & \colhead{DM - DM$_{\rm MW}^\ddag$} & \colhead{Redshift} & \colhead{Host Galaxy} & \colhead{Notes} & \colhead{$\log(M_\star)$} & \colhead{SFR} & \colhead{Localization } \\ 
\colhead{} & \colhead{[pc cm$^{-3}$]} & \colhead{} & \colhead{} & \colhead{} & \colhead{[M$_\odot$]} & \colhead{[M$_\odot$/{\rm yr}]} & \colhead{Reference(s)} } 

\startdata
20221012A & 386.68 & 0.284669 & PSO J280.8014+70.5242 & Quiescent & 10.99 & 0.15 & (1) \\
20220920A & 274.69 & 0.158239 & PSO J240.2568+70.9180 &  & 9.81 & 4.10 & (1) \\
20220914A & 576.08 & 0.113900 & PSO J282.0585+73.3363 &  & 9.48 & 0.72 & (1) \\
20220825A & 571.54 & 0.241397 & PSO J311.9820+72.5848 &  & 9.95 & 1.62 & (1) \\
20220610A & 1320$^\dagger$ & 1.017 &  & High-$z$ star-forming galaxy & 9.7 & 1.7 & (2) \\
20220509G & 214.33 & 0.089400 & PSO J282.6748+70.2427 & Spiral & 10.79 & 0.23 & (1) \\
20220506D & 307.87 & 0.30039 & PSO J318.0447+72.8272 &  & 10.29 & 5.08 & (1) \\
20220418A & 585.65 & 0.622000 & PSO J219.1065+70.0953 &  & 10.31 & 29.41 & (1) \\
20220310F & 416.84 & 0.477958 & PSO J134.7211+73.4910 &  & 9.97 & 4.25 & (1) \\
20220307B & 363.57 & 0.248123 & PSO J350.8747+72.1918  &  & 10.04 & 2.71 & (1) \\
20220207C & 183.08 & 0.043040 & PSO J310.1977+72.8826 & Star-forming disk galaxy & 9.91 & 1.16 & (1) \\
20211212A & 178.9 & 0.0715 &  & Star-forming spiral & 10.28 & 0.73 & (3), (4), (5) \\
20211203C & 572.8 & 0.3439 &  & Star-forming & 9.76 & 15.91 & (3), (4), (6) \\
20211127I & 192.33 & 0.0469 & WALLABY J131913-185018 & Star-forming spiral & 9.48 & 35.83 & (7), (5) \\
20210807D & 130.7 & 0.1293 &  & Quiescent spiral & 10.97 & 0.63 & (3), (5) \\
20210410D & 522.58 & 0.1415 & J214420.69-791904.8 & Star-forming & 9.47 & 0.03 & (8) \\
20210405I & 50.3 & 0.066 & J1701249-4932475 & Star-forming spiral & 11.25 & 0.3 & (9) \\
20210320C & 342.6 & 0.2797 &  & Star-forming & 10.37 & 3.51 & (3), (4), (6) \\
20210117A & 695.1 & 0.2145 & J223955.07-160905.37 & Dwarf galaxy & 8.59 & 0.02 & (10) \\
20201124A & 287.03 & 0.0979 & SDSS J050803.48+260338.0 & Massive dusty star-forming galaxy & 10.23 & 2.1 & (11), (12), (13) \\
20200906A & 541.8 & 0.3688 & PSO J033358.994-140459.287 & Star-forming galaxy & 10.12 & 0.48 & (14) \\
20200430A & 352.9 & 0.1608 & PSO J151849.52+122235.8 & No reported structure & 9.11 & 0.2 & (15) \\
20200223B & 156.2 & 0.06024 & SDSS J003304.68+284952.6 &  & 10.75 & 0.59 & (16) \\
20191228A & 264.5 & 0.2432 &  & Star-forming galaxy & 9.73 & 0.50 & (14) \\
20191106C & 307.2 & 0.10775 & SDSS J131819.23+425958.9 &  & 10.65 & 4.75 & (16) \\
20191001A & 463.9 & 0.2340 & DES J213324.44-544454.65 & Star-forming spiral, merger & 10.7 & 11.2 & (17), (18), (19)\\
20190714 & 466.13 & 0.2365 & PSO J121555.0941-130116.004 & Possible spiral arms & 10.17 & 0.65 & (15) \\
20190711A & 537.23 & 0.5217 &  & No structure & 8.91 & 0.42 & (15) \\
20190611 & 263.57 & 0.3778 & J212258.0-792350 & No spiral arms & 8.9 & 0.27 & (15) \\
20190608B & 301.5 & 0.1178 & SDSS J221604.90-075355.9 & Star-forming spiral & 10.4 & 1.2 & (20) \\
20190523A & 723.8 & 0.66 & DSA-10 J134815.6+722811 & Quiescent & 10.79 & $<$0.09 & (15) \\
20190520B & 642$^*$ & 0.241 & J160204.31-111718.5 & Dwarf galaxy & 8.78 & 0.41 & (21), (22) \\
20190425A & 79.2 & 0.03122 & UGC 10667 & Star-forming spiral & 10.26 & 1.5 & (23) \\
20190418A & 113.5 & 0.07132 & SDSS J042314.96+160425.6 & Star-forming spiral & 10.27 & 0.15 & (23) \\
20190102C & 306.3 & 0.291 & ASKAP-ICS J212939.76-792832.5 & Star-forming & 9.53 & 0.86 & (15) \\
20181223C & 92.5 & 0.03024 & SDSS J120340.98+273251.4 & Star-forming spiral & 9.29 & 0.054 & (23) \\
20181220A & 83.4 & 0.02746 & 2MFGC 17440 & Star-forming spiral & 9.86 & 3.0 & (23) \\
20181112A & 487.27 & 0.4755 & ASKAP-ICS J214923.63-525815.4 & Star-forming, no identified spiral & 9.42 & 0.6 & (20) \\
20181030A & 62.5 & 0.0039 & NGC 3252 & Star-forming barred spiral galaxy & 9.76 & 0.355 & (24) \\
20180924B & 320.92 & 0.3214 & ASKAP-ICS J214425.255-405400.1 & Star-forming & 10.35 & 0.88 & (15), (20) \\
20180916B & 149.06 & 0.0337 & SDSS J015800.28+654253.0 & Star-forming/Quiescent spiral & 9.33 & 0.06 & (15) \\
20180301A & 384 & 0.3304 & PSO J093.2268+04.6703 & Star-forming galaxy & 9.36 & 1.93 & (14) \\
20121102A & 372.58 & 0.19273 & J053158.698+330852.60 & Star-forming dwarf galaxy & 8.14 & 0.15 & (25) \\
\enddata

\tablenotetext{}{$^\ddag$Reported DM with MW contributions subtracted (usually \citealt{Cordes2002} model), $^\dagger$Intergroup-medium subtracted, $^*$Intercluster-media subtracted.}


\tablerefs{ (1) \citealt{Law2024}, (2) \citealt{Gordon2024}, (3) \citealt{Gordon2023}, (4) \citealt{James2022}, (5) Deller in prep., (6) Shannon in prep., (7) \citealt{Glowacki2023}, (8) \citealt{Caleb2023}, (9) \citealt{Driessen2024}, (10) \citealt{Bhandari2023}, (11) \citealt{Ravi2022}, (12) \citealt{Fong2021}, (13) \citealt{Piro2021}, (14) \citealt{Bhandari2022}, (15) \citealt{Heintz2020}, (16) \citealt{Ibik2024}, (17) \citealt{Bhandari2020b}, (18) \citealt{Kundu2022}, (19) \citealt{Woodland2023}, (20) \citealt{Bhandari2020a}, (21) \citealt{Niu2022}, (22) \citealt{Lee2023}, (23) \citealt{Bhardwaj2023}, (24) \citealt{Bhardwaj2021}, (25) \citealt{Tendulkar2017}. }
\label{table:frbs}
\end{deluxetable}

\rev{We have assembled, as far as the authors can tell, a near complete collection of the published FRBs with identified host galaxies and redshifts as of July 31, 2024 to compare against the host galaxy dispersion measure distributions found in the FIRE-2 galaxies.  The FRBs, their identifiers, extragalactic DMs, redshifts, host galaxy identifiers, stellar masses, star formation rates, and localization references can be found in Table~\ref{table:frbs}.  Two of the FRBs, 20190520B and 20220610A have additional reported DM contributions from galaxy clusters/groups along the line of sight, which we have subtracted according to their respective source publications. A large fraction of the reported host galaxies, where morphology has been noted do appear to be star-forming spirals (about a third of the sample, though by-and-large the whole sample of host galaxies has not been observed at high spatial resolution).  Figure~\ref{fig:hostgal_dist} shows both the inferred host galaxy DM distribution (by subtracting the mean IGM component, as described in \ref{sec:meth}), from which we derive our 5-95\% observations distribution that we compare against FIRE-2 in figures in the main text, and the extragalactic DM excess--redshift (the `\citealt{2020Natur.581..391M}' relation) showing the derived 5-95\% range of inferred ${\rm DM_{Host}}$.  }

\section{High-redshift Spatially Resolved Dispersion Measure Maps}\label{sec:append:DMmaps}

\begin{figure*}
    \centering
    \includegraphics[width=0.9\linewidth]{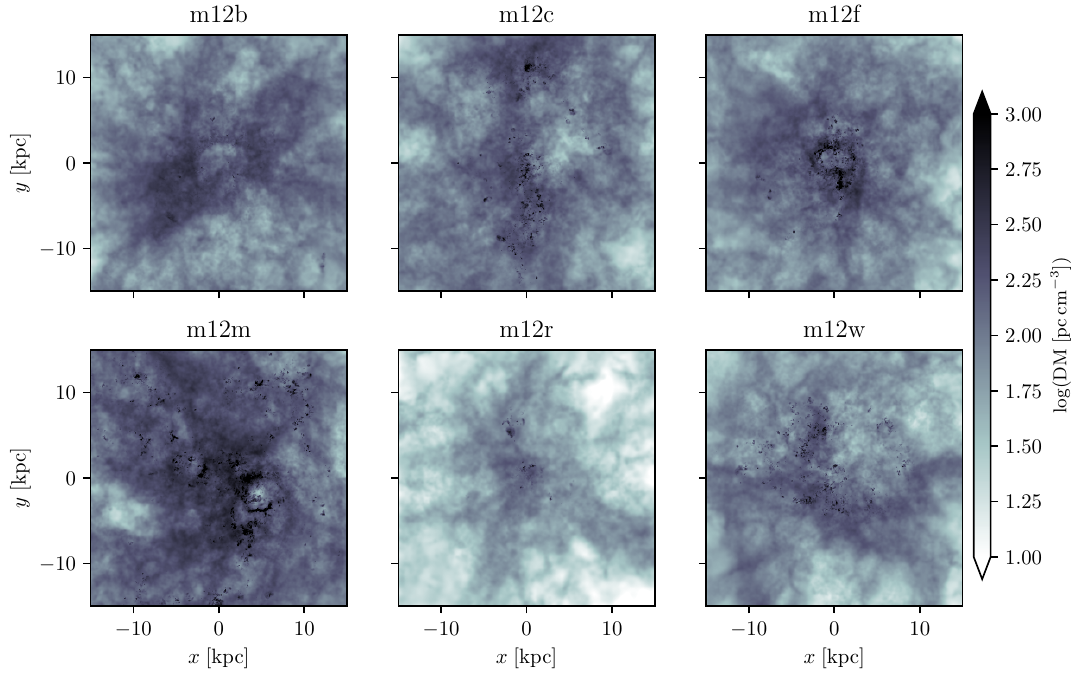}
    \caption{Spatially resolved (100 pc smoothing) face-on maps of dispersion measure (DM) in the six FIRE-2 simulated galaxies analyzed in this paper, at $z=1$, in the style of Fig.~\ref{fig:z0maps}. All six galaxies are highly irregular, hosting bursty star formation and disturbed gas reservoirs with high DMs.}
    \label{fig:z1maps}
\end{figure*}

\begin{figure*}
    \centering
    \includegraphics[width=0.9\linewidth]{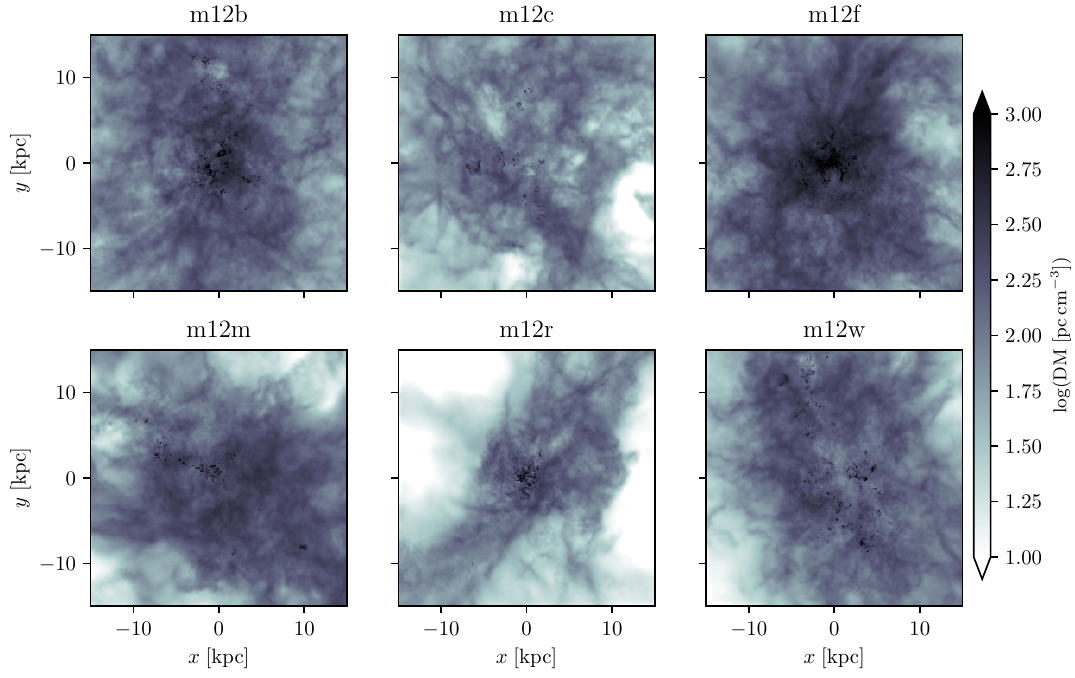}
    \caption{Spatially resolved (100 pc smoothing) face-on maps of dispersion measure (DM) in the six FIRE-2 simulated galaxies analyzed in this paper, at $z=2$, in the style of Fig.~\ref{fig:z0maps}. All six galaxies are highly irregular, hosting bursty star formation and disturbed gas reservoirs with high DMs.}
    \label{fig:z2maps}
\end{figure*}

We present additional dispersion measure (DM) maps at 100~pc spatial resolution of the full free-electron column of each of the six FIRE-2 galaxies analyzed in this paper at $z=1$ and $z=2$, in the style of Figure~\ref{fig:z0maps}. At both redshifts, all of the simulated galaxies are highly irregular, dispersion-supported objects.  Highly clustered star formation is visible in many of the images as tight knots of high-DM H{\scriptsize II} regions.  A few of the images, \textbf{m12b} and \textbf{m12r} at $z=1$ and \textbf{12c} at $z=2$, show the galaxies in somewhat post-starburst states with them having little to no star formation in the past $\sim$Myr.  These simulated galaxies are missing the smattering of very high DM very young H{\scriptsize II} regions seen in the other runs (\textit{cf.} Figures~\ref{fig:z1dists} \& \ref{fig:z2dists}).

\section{Stacked Distribution of Dispersion Measure from FIRE Galaxy Sample at Each Redshift}\label{sec:append:stacks}
\begin{figure*}
	\includegraphics[width=\textwidth]{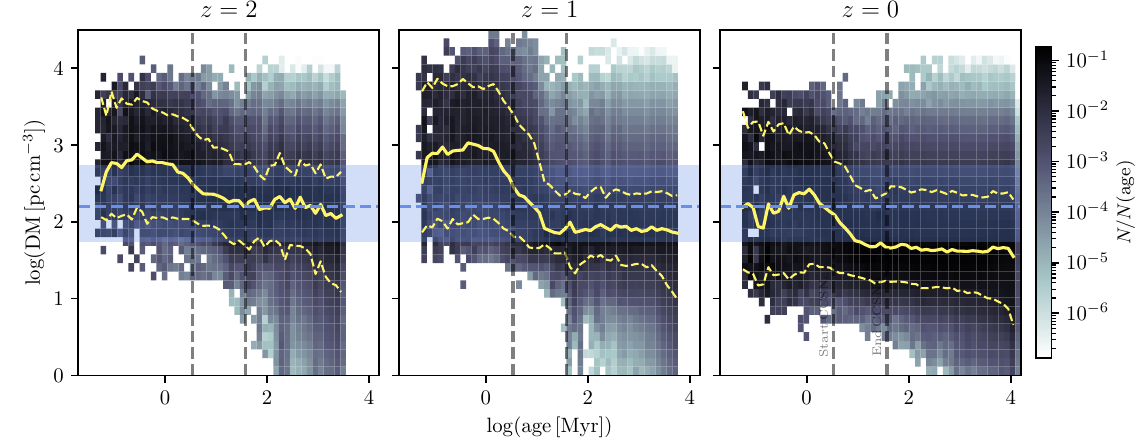}
    \caption{Distributions of dispersion measure (DM, \ie free electron column density) of all six Milky Way mass FIRE-2 simulated galaxies analyzed in this letter at $z=2-0$ stacked together, style as Fig.~\ref{fig:z0dists}. DM distribution with age has a similar shape through cosmic time, with the youngest stars being embedded in highly ionized HII regions, and evolving into the less ionized ambient ISM- albeit with a high degree of scatter due to chance alignments with HII regions or having little to no galactic ISM along the line of sight.}
    \label{fig:z20stackeddists}
\end{figure*}

In Figure~\ref{fig:z20stackeddists} we stack together all six of the FIRE-2 galaxies analyzed here to show compactly how the total distribution of DM versus stellar age evolves with redshift. DM--stellar age distributions in the FIRE galaxies generally have the same form across redshift, and the changing morphology of the galaxies.  However, as the galaxies evolve from non-disky bursty galaxies to disks at redshift zero, the DMs fall as the specific star formation rates reduce and the ISM cools (and the overall ionization fraction falls).

\section{Individual Galaxy Dispersion Measure 1-D Distributions}\label{sec:append:PDFs}
\begin{figure*}
	\includegraphics[width=\textwidth]{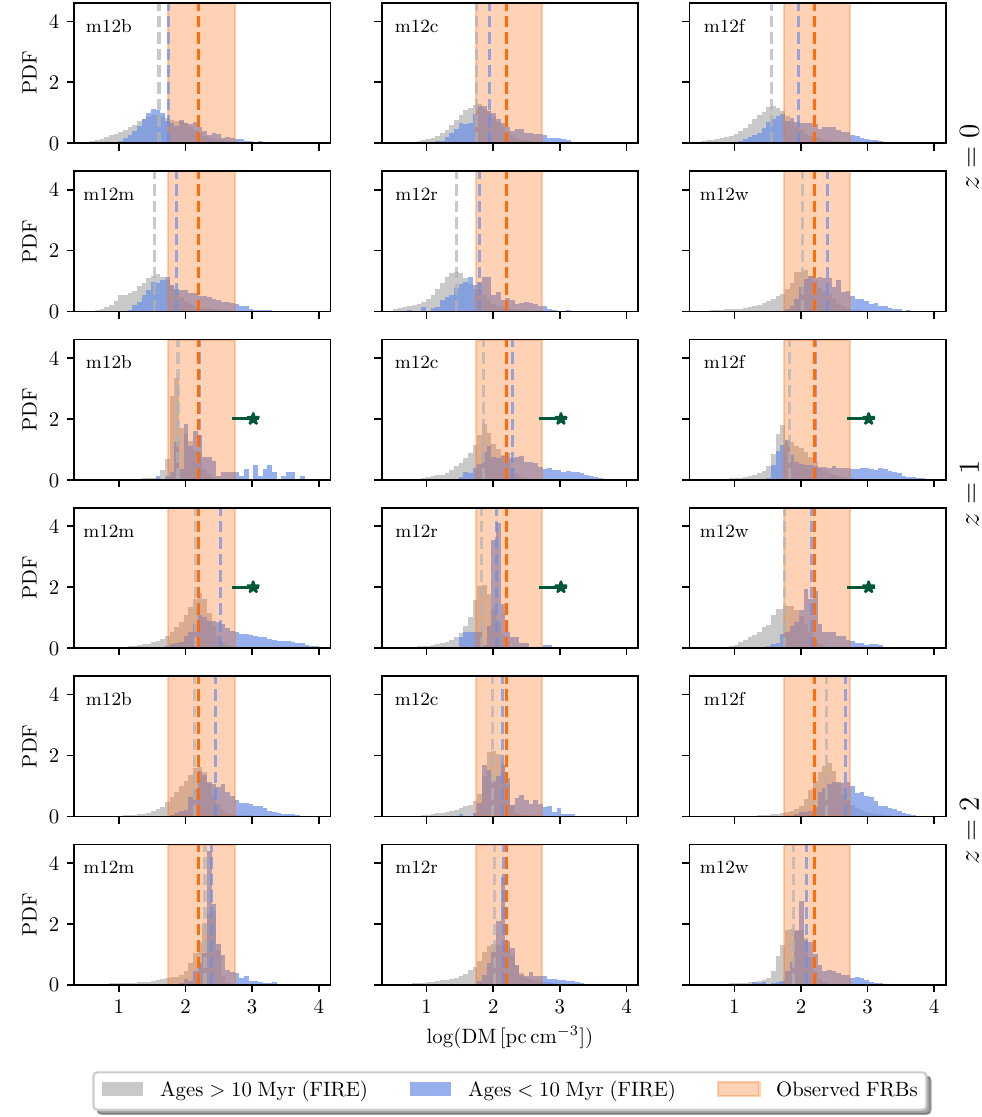}
    \caption{Individual galaxy distributions of host-galaxy dispersion measure for all stars in the FIRE-2 simulations at $z = 0,1,$ \& 2, in the style of Fig.~\ref{fig:stackedDMPDFs}. Significant galaxy-to-galaxy variation in DM distribution (and between young and old stellar populations) is seen at all redshift. In many cases, observed FRB DMs may be entirely accounted for by host-galaxy DM contributions in the simulations, especially if FRB sources are young stars.}
    \label{fig:zPDFs}
\end{figure*}

Here we show DM distributions from all six FIRE-2 galaxies analyzed in this paper at $z = 0,1$ \& 2, in Figure~\ref{fig:zPDFs}.  These are all the underlying distributions that are stacked together in Figure~\ref{fig:stackedDMPDFs}.  Significant galaxy-to-galaxy variation is seen at all redshift, and between older and younger stellar populations within galaxies.  A large fraction of observed FRB DMs can be explained entirely by the host-galaxy contribution, if FRBs are sourced by a young stellar population, or at higher redshift by an older stellar population.

\end{document}